\documentclass[aps,pre,twocolumn,tightenlines,graphicx]{revtex4}

\usepackage[dvips]{graphicx}
\newcommand{\la}{\left\langle}
\newcommand{\ra}{\right\rangle}
\newcommand{\EPL}{{\it Europhys.~Lett.~}}
\newcommand{\PRL}{{\it Phys.~Rev.~Lett.~}}
\newcommand{\PR}{{\it Phys.~Rev.~}}
\newcommand{\JCP}{{\it J.~Chem.~Phys.~}}
\newcommand{\JPCM}{{\it J.~Phys.: Condens.~Matter~}}
\newcommand{\MP}{{\it Mol.~Phys.~}}

\begin{document}


\title{Counterion Penetration and Effective Electrostatic Interactions
in Solutions of Polyelectrolyte Stars and Microgels}


\author{A. R. Denton}
\email{alan.denton@ndsu.nodak.edu}
\affiliation{Department of Physics, North Dakota State University,
Fargo, ND, 58105-5566}


\date{\today}

\begin{abstract}
Counterion distributions and effective electrostatic interactions
between spherical macroions in polyelectrolyte solutions are
calculated via second-order perturbation (linear response) theory.
By modelling the macroions as continuous charge distributions that
are permeable to counterions, analytical expressions are obtained
for counterion profiles and effective pair interactions in
solutions of star-branched and microgel macroions.
The counterions are found to penetrate stars more easily than
microgels, with important implications for screening of bare
macroion interactions. The effective pair interactions are Yukawa 
in form for separated macroions, but are softly repulsive and bounded 
for overlapping macroions.  A one-body volume energy, which depends 
on the average macroion concentration, emerges naturally in the theory 
and contributes to the total free energy.
\end{abstract}

\pacs{82.70.Dd, 82.45.-h, 05.20.Jj}

\maketitle

\section{Introduction}

Polyelectrolytes (PEs) are ionizable polymers that dissolve in a polar
solvent, such as water, by dissociating into polyvalent macroions and
small oppositely-charged counterions~\cite{PE1}.  Electrostatic
interactions between macroions, mediated by surrounding microions
(counterions and salt ions), contribute to the unique macroscopic
properties of PE solutions, which are the basis of many industrial
applications involving polymer-water systems~\cite{PE2}.
Common synthetic examples of PEs are polyacrylic acid,
used in gels and rheology modifiers, and polystyrene sulfonate, a
component of reverse osmosis membranes.  Naturally occurring examples
are biopolymers, such as DNA, proteins, and starches.
Colloidal in size, PEs are also routinely added as flocculants and
stabilizers to colloidal suspensions, such as foods and water-based
paints~\cite{Evans,Hunter}.  Depending on PE concentration,
adsorption or grafting of PE chains onto surfaces of colloidal particles
can either induce flocculation, by bridging particles, or impart
electrosteric stabilization.

Conformations of PE macroions and electrostatic interactions between
macroions are strongly influenced by the distribution of microions.
If dispersed in solution, microions act to screen the bare Coulomb
interactions between ionized monomers.  If condensed on the macroion
chains, microions may reduce the macroion charge~\cite{Schiessel}.
Linear PE chains whose monomers are sufficiently weakly interacting
-- either because weakly charged or because of strong microion screening
or condensation -- may form random-walk coils with roughly spherical
conformation.  With increasing charge and
screening length, linear chains stretch into non-spherical
conformations because of electrostatic repulsion between ionized
monomers~\cite{deGennes,Khokhlov}.  The extent of elongation
depends on the chain charge density, salt concentration, and
solvent quality. Highly charged chains in good solvents ({\it e.g.}, 
DNA in water) often form stiff rod-like macroions, whose
effective interactions and complex phase behavior (such as
bundling) have been widely studied~\cite{rods1,rods2}. In poor solvents,
sufficiently highly charged chains may form necklaces of compact
globules joined by narrow threads, as predicted by
theory~\cite{Rubinstein} and confirmed by
simulation~\cite{Kremer}.

Although many common PEs are linear, other topologies can be
readily synthesized.  Examples are stars, microgels, micelles, and
brushes. Star polymers~\cite{Likos1} consist of chains chemically
grafted or adsorbed to a common microscopic core.  Microgels are
mesoscopic polymer networks, synthesized by polymerization in
microemulsion~\cite{microgels}. Micelles are formed by association
of charged diblock (amphiphilic) copolymers~\cite{micelles}.
Brushes are formed by grafting PE chains onto a mesoscopic
solid core~\cite{brushes}. Solutions of spherical stars, microgels,
micelles, and brushes can be regarded as colloidal suspensions
of soft macroions that are permeable to microions.

Electrostatic interactions in charged colloids have received much
attention in recent years~\cite{reviews}, motivated largely by
anomalous phase behavior that is unexplained by the classic 
Derjaguin-Landau-Verwey-Overbeek (DLVO) theory~\cite{DLVO}.
Most studies have been restricted, however, to hard, impermeable
macroions. The objectives of this paper are first, to explore
implications of microion penetration for screening of effective
electrostatic interactions between spherical macroions, and
second, to lay a foundation for future studies of thermodynamic
phase behavior of PE solutions. Our approach is based on a
recently-proposed theory of effective interactions in charged
colloids~\cite{Denton1}, which we adapt here from hard to
penetrable macroions and apply to spherical star-branched and
microgel macroions.

The remainder of the paper is organized as follows.
Section~\ref{Model} describes the assumed model of PE solutions.
Section~\ref{Theory} reviews the theoretical approach, based on
second-order perturbation (linear response) theory.
Sections~\ref{Analytical Results} and \ref{Numerical Results}
present analytical and numerical results for counterion profiles
and effective interactions in bulk solutions of star and microgel
macroions. Finally, Sec.~\ref{Conclusions} closes with a summary
and conclusions.

\section{Model}\label{Model}
Adapting the primitive model of ionic liquids~\cite{HM}, the model
system comprises $N_m$ spherical macroions, of radius $a$ 
(diameter $\sigma=2a$) and charge $-Ze$, and $N_c$ point counterions of
charge $ze$ dispersed in an electrolyte solvent in a volume $V$
at temperature $T$. Assuming, for simplicity, a symmetric
electrolyte and equal salt and counterion valences, the
electrolyte contains $N_s$ point salt ions of charge $ze$ and
$N_s$ of charge $-ze$. The microions thus number $N_+=N_c+N_s$
positive and $N_-=N_s$ negative, for a total of
$N_{\mu}=N_c+2N_s$. Global charge neutrality in a bulk solution
constrains average macroion and counterion number densities, 
$n_m=N_m/V$ and $n_c=N_c/V$, via $Zn_m=zn_c$. The polar solvent 
is treated as a continuum, characterized by dielectric constant 
$\epsilon$ that acts only to reduce Coulomb interactions between ions.

The local number density profiles of counterions, $\rho_c(r)$, 
and of macroion monomers, $\rho_{\rm mon}(r)$, are modelled as 
spherically symmetric, continuous distributions.  Spherical symmetry 
is a reasonable approximation, considering that equilibrium averaging
over macroion orientations tends to smear out any anisotropy.
Furthermore, discreteness of the charge distributions can be ignored 
if we restrict consideration to length scales exceeding the 
scale of discreteness.

In general, the counterions are distributed over three regions:
(1) the immediate vicinity of the PE chains making up the macroions,
(2) the region inside of the macroions but away from the chains, and
(3) the region outside of the macroions.
Counterions in the first two regions are trapped by the macroions,
while those in the third region are free.
Within the first region, the counterions may be either condensed 
on a chain or free to move along a tube surrounding a chain.  
These chain-localized counterions, whether condensed or mobile, 
tend to distribute uniformly along the chains to favor local charge 
neutrality.  In our model, counterions in region (1) simply 
renormalize the effective macroion valence $Z$.  

The detailed form of the monomer density profile depends on 
the macroion conformation.
For star-branched macroions, Coulomb repulsion between charged
monomers tends to stiffen and radially stretch the chains into a
porcupine conformation~\cite{Pincus}. We assume the ideal case 
of fully stretched chains and model the monomer density profile by
$\rho_{\rm mon}(r\leq a)=Z/(4\pi ar^2)$,
where $r$ is the radial distance from the star's center. For
microgel macroions, the dense network of chains is well
approximated by a uniform monomer distribution, and is modelled
here by $\rho_{\rm mon}(r\leq a)=3Z/(4\pi a^3)$. This distribution may
also approximate a weakly-charged linear PE chain with a spherical
random-coil conformation, although a Gaussian distribution may then
be more accurate. For both the star and microgel models, the monomer
density profile is cut off sharply at the macroion surface:
$\rho_{\rm mon}(r>a)=0$.

\section{Theory}\label{Theory}

For the model PE solutions described above, the theoretical
challenge is to predict the distributions of microions inside and
outside of the macroions and the effective interactions between
macroions. Following the same general strategy as applied
previously to charged colloids~\cite{Denton1,Silbert}, we reduce
the multi-component mixture to an equivalent one-component system
governed by effective interactions, which are approximated via
perturbation theory. For clarity of presentation, we initially
ignore salt ions. The Hamiltonian then decomposes into three
terms:
\begin{equation}
H=H_m(\{{\bf R}\})+H_c(\{{\bf
r}\})+H_{mc}(\{{\bf R}\},\{{\bf r}\}), \label{H}
\end{equation}
where $\{{\bf R}\}$ and $\{{\bf r}\}$ denote collective
coordinates of macroion centers and counterions, respectively.
The first term,
\begin{equation}
H_m=K_m+\frac{1}{2}\sum_{i\neq j=1}^{N_m}v_{mm}(|{\bf R}_i-{\bf
R}_j|), \label{Hp}
\end{equation}
is the bare Hamiltonian for macroions with kinetic energy $K_m$ that
interact via the bare pair potential $v_{mm}(r)$ at center-center
separation $r$.  The form of $v_{mm}(r)$ depends on the macroion
conformation and is specified in the Appendix.
The second term in Eq.~(\ref{H}),
\begin{equation}
H_c=K_c+\frac{1}{2}\sum_{i\neq j=1}^{N_c}v_{cc}(|{\bf r}_i-{\bf
r}_j|), \label{Hc}
\end{equation}
is the Hamiltonian for counterions with kinetic energy $K_c$ that
interact via the Coulomb pair potential
$v_{cc}(r)=z^2e^2/\epsilon r$. The third term in Eq.~(\ref{H}),
\begin{equation}
H_{mc}=\sum_{i=1}^{N_m}\sum_{j=1}^{N_c}v_{mc}(|{\bf R}_i-{\bf
r}_j|), \label{Hmc1}
\end{equation}
is the macroion-counterion interaction. For spherical macroions,
\begin{equation}
v_{mc}(r)~=~\left\{ \begin{array} {l@{\quad\quad}l}
\frac{\displaystyle -Zze^2}{\displaystyle \epsilon r}, & r>a \\
v_<(r), & r\leq a,
\end{array} \right.
\label{vmcr}
\end{equation}
where the interaction inside a macroion, $v_<(r)$, depends on the
macroion conformation and is specified in Sec.~\ref{Analytical Results}.
For later reference, we note that
Eq.~(\ref{Hmc1}) also may be expressed in the form
\begin{equation}
H_{mc}~=~\int{\rm d}{\bf R}\,\rho_m({\bf R})\int{\rm d}{\bf
r}\,\rho_c({\bf r})v_{mc}(|{\bf R}-{\bf r}|), \label{Hmc2}
\end{equation}
where $\rho_m({\bf R})=\sum_{j=1}^{N_m} \delta({\bf R}-{\bf R}_j)$
and $\rho_c({\bf r})=\sum_{j=1}^{N_c} \delta({\bf r}-{\bf r}_j)$
are the macroion and counterion number density operators, respectively.

The mixture of macroions and counterions is formally reduced to an
equivalent one-component system by tracing over counterion
coordinates. Denoting counterion and macroion (classical) traces
by $\la~\ra_c$ and $\la~\ra_m$, respectively, the canonical
partition function can be expressed as
\begin{equation}
{\cal Z}~=~\la\la\exp(-\beta H)\ra_c\ra_m~=~\la\exp(-\beta H_{\rm
eff})\ra_m, \label{part}
\end{equation}
where $H_{\rm eff}=H_m+F_c$ is the effective one-component
Hamiltonian, $\beta=1/k_BT$, and
\begin{equation}
F_c~=~-k_BT\ln\la\exp\Bigl[-\beta(H_c+H_{mc})\Bigr]\ra_c
\label{Fc1}
\end{equation}
is the free energy of a nonuniform gas of counterions in the
presence of the macroions. 

At this stage, approximations are necessary for the counterion free energy.
It is first convenient to convert the counterion Hamiltonian to 
the Hamiltonian of a classical one-component plasma (OCP) of 
counterions by adding to $H_c$, and subtracting from $H_{mc}$, 
the energy of a uniform compensating negative background~\cite{note1}, 
$E_b=-N_cn_c\hat v_{cc}(0)/2$, where $\hat v_{cc}(0)$ is the 
$k \to 0$ limit of the Fourier transform of $v_{cc}(r)$. 
Now regarding the macroions as an ``external" potential for the OCP, 
we invoke perturbation theory~\cite{HM,Denton1,Silbert} and write
\begin{equation}
F_c~=~F_{\rm OCP}~+~\int_0^1{\rm d}\lambda\,\la
H'_{mc}\ra_{\lambda}, \label{Fc2}
\end{equation}
where $F_{\rm OCP}=-k_BT\ln\la\exp[-\beta(H_c+E_b)]\ra_c$ is the
OCP free energy, the $\lambda$-integral charges the macroions,
$H'_{mc}=H_{mc}-E_b$ represents the perturbing potential of the
macroions acting on the counterions, and $\la H'_{mc}\ra_{\lambda}$
is the mean value of this potential in a solution of macroions 
charged to a fraction $\lambda$ of their full charge. 
Further progress is facilitated by expressing $H_{mc}$ [Eq.~(\ref{Hmc2})] 
in terms of Fourier components:
\begin{eqnarray}
&&\la H_{mc}\ra_{\lambda}=\frac{1}{V}\sum_{{\bf k}\neq 0} \hat
v_{mc}(k) \hat\rho_m(-{\bf k}) \la\hat\rho_c({\bf k})\ra_{\lambda} \nonumber \\
~~&+&\frac{1}{V}\lim_{k\to 0}\left[\hat v_{mc}(k) \hat\rho_m(-{\bf
k})\la\hat\rho_c({\bf k})\ra_{\lambda}\right], \label{Hmck}
\end{eqnarray}
where $\hat v_{mc}(k)$ is the Fourier transform of
Eq.~(\ref{vmcr}) and where $\hat\rho_m({\bf
k})=\sum_{j=1}^{N_m}\exp(i{\bf k}\cdot{\bf R}_j)$ and
$\hat\rho_c({\bf k})=\sum_{j=1}^{N_c}\exp(i{\bf k}\cdot{\bf r}_j)$
are Fourier components of the macroion and counterion densities.

In first-order perturbation theory, the response of the counterion
plasma to the macroions is ignored.  Here we apply second-order
perturbation (linear response) theory, in which the counterions
are assumed to respond linearly to the macroion external potential:
\begin{equation}
\hat\rho_c({\bf k})~=~\chi(k)\hat v_{mc}(k)\hat\rho_m({\bf k}),
\qquad k\neq 0, \label{rhock}
\end{equation}
where $\chi(k)$ is the linear response function of the OCP. Note
that the $k\to 0$ limit here, and in Eq.~(\ref{Hmck}), must be
treated separately, since the average counterion density,
$n_c=\hat\rho_c(0)$, does not respond to the macroion charge, but
rather is fixed by the constraint of global charge neutrality.

Upon combining Eqs.~(\ref{Fc2})-(\ref{rhock}), the effective
Hamiltonian can be recast in the form of the Hamiltonian of a
pairwise-interacting system:
\begin{equation}
H_{\rm eff}~=~K_m~+~K_c~+~\frac{1}{2}\sum_{i\neq j=1}^{N_m}v_{\rm
eff}(|{\bf R}_i-{\bf R}_j|)~+~E_0, \label{Heff}
\end{equation}
where $v_{\rm eff}(r)=v_{mm}(r)+v_{\rm ind}(r)$ is an effective
macroion pair interaction that combines the bare macroion
interaction with a microion-induced interaction
\begin{equation}
\hat v_{\rm ind}(k)~=~\chi(k)\left[\hat v_{mc}(k)\right]^2.
\label{vindk}
\end{equation}
The final term in Eq.~(\ref{Heff}) is the volume energy, formally
given by
\begin{eqnarray}
&&E_0=F_{\rm OCP}+\frac{N_m}{2}\lim_{r\to 0} v_{\rm ind}(r) \nonumber \\
&+&N_m\lim_{k\to 0}\left[-\frac{n_m}{2}\hat v_{\rm ind}(k) 
+n_c\hat v_{mc}(k)+\frac{Zn_c}{2z}\hat v_{cc}(k)\right], \nonumber \\
\label{E0}
\end{eqnarray}
which is a natural by-product of the one-component reduction.
Although independent of the macroion coordinates, the volume
energy depends on the average macroion density and thus can
influence thermodynamics.

The linear response function is proportional to the corresponding
static structure factor, $S(k)$, which may be obtained from
liquid-state theory~\cite{HM}.  
In practice, the OCP is weakly correlated, with coupling parameter 
$\Gamma=\lambda_B/a_c \ll 1$, where
$\lambda_B=\beta e^2/\epsilon$ is the Bjerrum length and
$a_c=(3/4\pi n_c)^{1/3}$ is the counterion sphere radius.
For example, for macroions of diameter $\sigma=100$ nm, valence $Z=100$, 
and volume fraction $\eta=(\pi/6)n_m\sigma^3=0.01$, 
in water at room temperature ($\lambda_B=0.714$ nm), we find 
$\Gamma\simeq 0.014$.
As for charged colloids~\cite{Denton1,Silbert}, we adopt the random phase
approximation (RPA), which is accurate for weakly-coupled plasmas.
The RPA equates the two-particle direct correlation function of
the OCP to its exact asymptotic limit: $c^{(2)}(r)=-\beta
v_{cc}(r)$. Using the Ornstein-Zernike relation,
$S(k)=1/[1-n_c\hat c^{(2)}(k)]$, the linear response function then
takes the analytical form
\begin{equation}
\chi(k)~=~-\beta n_cS(k)~=~-\frac{\beta n_c}{(1+\kappa^2/k^2)},
\label{chi}
\end{equation}
where $\kappa=\sqrt{4\pi n_cz^2\lambda_B}$ is the inverse Debye
screening length.  Note that since permeable macroions do not
exclude counterions from their interiors, the excluded-volume
corrections required for hard colloidal macroions~\cite{Denton1}
are not relevant here.  With $\chi(k)$ specified, the counterion
density can be explicitly determined from Eqs.~(\ref{vmcr}) and
(\ref{rhock}) for a given macroion distribution (see
Sec.~\ref{Analytical Results}). Finally, salt is easily introduced
via additional microion response functions. In the process, the
pair interaction and volume energy are unchanged, except for a
redefinition of the screening constant as $\kappa=\sqrt{4\pi
(n_c+2n_s)z^2\lambda_B}$, where $n_s$ is the average number density
of salt ion pairs.

It is worth noting the formal equivalence of the present theory to
linearized Poisson-Boltzmann (DLVO) theory.  Both are mean-field theories
in the sense that they ignore fluctuations in microion distributions. 
An advantage of linear response theory, however, is that it 
encompasses the volume energy, which can be important for
describing phase behavior~\cite{Denton1,Silbert,vRH,Graf,vRDH,Warren}.
Moreover, response theory can be
straightforwardly generalized to incorporate nonlinear response,
which entails both many-body effective interactions and
corrections to the pair potential and volume
energy~\cite{Denton2}. In contrast, nonlinear Poisson-Boltzmann
theory is practical only for the simple boundary conditions
afforded by cell models. For simplicity, higher-order nonlinear
effects are here ignored.

Equations~(\ref{rhock})-(\ref{E0}) constitute the main formal
expressions from linear response theory.  Explicit calculations 
require specifying the counterion-macroion interaction $v_<(r)$
in Eq.~(\ref{vmcr}) for specific macroion models.  Below, we apply
the theory to obtain analytical and numerical results for
counterion profiles and effective interactions in bulk solutions
of spherical star-branched and microgel macroions.

\section{Analytical Results}\label{Analytical Results}

\subsection{Star Macroions}\label{ResultsA}
For our idealized model of a star-branched macroion with $1/r^2$
monomer density profile, Gauss's law gives the electric field as
\begin{equation}
E(r)~=~\left\{ \begin{array}
{l@{\quad\quad}l}
-\frac{\displaystyle Ze}{\displaystyle \epsilon r^2}, & r>a \\
-\frac{\displaystyle Ze}{\displaystyle \epsilon ar}, & r\leq a.
\end{array} \right. \label{Estar}
\end{equation}
Integration over $r$ yields the electrostatic potential energy
between a star and a counterion:
\begin{equation}
v_{mc}(r)~=~\left\{ \begin{array} {l@{\quad\quad}l}
-\frac{\displaystyle Zze^2}{\displaystyle \epsilon r}, & r>a \\
-\frac{\displaystyle Zze^2}{\displaystyle \epsilon
a}~\left[1-\ln\left(\frac{\displaystyle r}{\displaystyle
a}\right)\right], & r\leq a,
\end{array} \right. \label{vmcrstar}
\end{equation}
whose Fourier transform is
\begin{equation}
\hat v_{mc}(k)~=~
-\frac{4\pi Zze^2}{\epsilon k^3a}~{\rm sinc}(ka), \label{vmckstar}
\end{equation}
with ${\rm sinc}(x)\equiv\int_0^x{\rm d}u\,\sin(u)/u$.
We can now calculate the counterion number density around a single
macroion in the dilute limit, where $\hat\rho_m({\bf k})=1$. From
Eqs.~(\ref{rhock}), (\ref{chi}), and (\ref{vmckstar}), the
Fourier component of the counterion density profile is given by
\begin{equation}
\hat\rho_c(k)~=~\frac{Z}{z}~\frac{\kappa^2}{ka(k^2+\kappa^2)}~{\rm
sinc}(ka), \label{rhockstar}
\end{equation}
whose real-space form is
\begin{equation}
\rho_c(r)~=~\frac{Z}{z}\frac{\kappa}{4\pi ar} 
~{\rm sinhc}(\kappa a)~e^{-\kappa r}, \quad r>a \label{rhocrstar-r>a}
\end{equation}
\begin{eqnarray}
\rho_c(r)~&=&~\frac{Z}{z}\frac{\kappa}{8\pi ar} 
\left[{\rm Ec}(\kappa a,\kappa r)+2~{\rm sinhc}(\kappa
a)\right]~e^{-\kappa r} \nonumber \\ 
~&-&~{\rm Ec}(-\kappa a,-\kappa r)~e^{\kappa r}, \quad r\leq a,
\label{rhocrstar-r<a}
\end{eqnarray}
where
\begin{eqnarray}
{\rm sinhc}(x)~&\equiv&~\int_0^x{\rm d}u\,\frac{\sinh(u)}{u} \nonumber \\
~&=&~\sum_{n=0}^{\infty}\frac{x^{2n+1}}{(2n+1)\cdot (2n+1)!}
\end{eqnarray}
and
\begin{equation}
{\rm Ec}(x_1,x_2)~\equiv~\int_{x_1}^{x_2}{\rm d}u\,\frac{e^u}{u}
~=~\ln\left(\frac{x_2}{x_1}\right)+\sum_{n=1}^{\infty}
\frac{x_2^n-x_1^n}{n\cdot n!},
\end{equation}
which can be efficiently computed from the first few terms
of the rapidly converging series expansions.
Approaching the macroion center, the counterion density profile 
varies more gradually than the $1/r^2$ macroion monomer density profile,
diverging logarithmically, according to
\begin{equation}
\lim_{r\to 0} \rho_c(r)~=~\frac{Z\kappa^2}{4\pi a}\left[1-\ln
\left(\frac{r}{a}\right)\right].
\end{equation}
Integrating Eq.~(\ref{rhocrstar-r<a}) over the spherical volume of 
the macroion yields the fraction of counterions inside a star:
\begin{eqnarray}
f_{\rm in}~&=&~\frac{z}{Z}4\pi\int_0^a{\rm d}r\,r^2\rho_c(r) \nonumber \\
~&=&~1-\left(1+\frac{1}{\kappa a }\right)~e^{-\kappa a}
~{\rm sinhc}(\kappa a). \label{ncinside-star}
\end{eqnarray}
Note the clear predictions that (1) the counterion distribution is
determined entirely by $\kappa a$, or the dimensionless ratio of the 
macroion radius and the Debye screening length, and (2) the fraction 
of counterions inside increases monotonically with $\kappa a$.  
Thus, for fixed macroion radius, $f_{\rm in}$ increases with 
increasing macroion valence and concentration.  This result is 
physically sensible: the shorter the screening length, the shorter 
the range of the counterion response, and thus the tighter the 
localization of counterions around the macroion centers.

From Eqs.~(\ref{vindk}) and (\ref{vmckstar}), the induced
electrostatic pair interaction is given by
\begin{equation}
\hat v_{\rm ind}(k) ~=~-\frac{4\pi
Z^2e^2}{\epsilon}~\frac{\kappa^2}{k^4a^2(k^2+\kappa^2)}~{\rm
sinc}^2(ka). \label{vindk-star}
\end{equation}
Fourier transforming, we obtain
\begin{eqnarray}
v_{\rm ind}(r)&=&-\frac{16\pi^2Z^2e^2\kappa^2a^2}{\epsilon r
}~\int_0^{\infty}{\rm
d}x\,\frac{\sin(xr/a)}{x^3(x^2+\kappa^2a^2)} \nonumber \\
~&\times&~{\rm sinc}^2x.
\label{vindr-star}
\end{eqnarray}
For nonoverlapping stars, Eq.~(\ref{vindr-star}) can be reduced to
the analytical form
\begin{equation}
v_{\rm ind}(r>2a)~=~-\frac{Z^2e^2}{\epsilon
r}~+~\frac{Z^2e^2}{\epsilon}~\left[\frac{{\rm sinhc}(\kappa a)}
{\kappa a}\right]^2~\frac{e^{-\kappa r}}{r}.
\label{vindr>2a-star}
\end{equation}
Since nonoverlapping macroions interact via a bare Coulomb
potential, $v_{mm}(r)=Z^2e^2/\epsilon r$, the effective pair
interaction for this case is
\begin{equation}
v_{\rm eff}(r>2a)~=~\frac{Z^2e^2}{\epsilon}~\left[\frac{{\rm sinhc}
(\kappa a)}{\kappa a}\right]^2~\frac{e^{-\kappa r}}{r}.
\label{veffr>2a-star}
\end{equation}
Thus, at the level of linear response, nonoverlapping star macroions 
interact via an effective Yukawa (screened-Coulomb) pair potential.  
The screening constant, $\kappa$, in the potential depends on the 
total density of microions -- inside and outside of the macroions -- 
since all microions respond to the macroion charge.
Note that the potential has the same $r$-dependence as 
the DLVO potential for hard colloidal macroions~\cite{Hunter,DLVO}, 
\begin{equation}
v_{\rm DLVO}(r)~=~\frac{Z^2e^2}{\epsilon}~\left[\frac{\exp(\kappa
a)}{1+\kappa a}\right]^2~\frac{e^{-\kappa r}}{r}, \qquad r>2a,
\label{vdlvo}
\end{equation}
differing only in the macroion-size-dependent amplitude. 
For overlapping stars, the bare macroion interaction is 
somewhat more complex and is relegated to the Appendix.

Finally, from Eqs.~(\ref{E0}), (\ref{vmckstar}),
(\ref{vindk-star}), and (\ref{vindr-star}), the volume energy is
obtained as
\begin{equation}
E_0=F_{\rm OCP}-N_m\frac{8\pi^2Z^2e^2\kappa^2a}{\epsilon}
\int_0^{\infty}{\rm d}x\,\frac{{\rm
sinc}^2x}{x^2(x^2+\kappa^2a^2)}. \label{E0-star}
\end{equation}
For weakly-coupled microion plasmas, the OCP free energy may be
approximated by its ideal-gas limit:
\begin{equation}
F_{\rm OCP}~=~N_+[\ln(n_+\Lambda^3)-1]~+~N_-[\ln(n_-\Lambda^3)-1],
\label{FOCP}
\end{equation}
where $\Lambda$ is the thermal wavelength. Note that if $Z$ is
allowed to vary (with counterion condensation), then the volume
energy per macroion must be augmented by the self (Hartree) energy
of a macroion: $U_H=Z^2e^2/\epsilon a$.
The first term in Eq.~(\ref{E0-star}), long recognized as
important for phase behavior~\cite{PE1}, represents the entropy of
free counterions; the second term accounts for the cohesive
electrostatic energy of microion-macroion interactions.  
The volume energy, analogous to its counterpart for charged
colloids~\cite{Denton1,vRDH,Graf,vRDH}, depends on the average macroion 
concentration and thus has the potential to influence phase behavior 
and other thermodynamic properties.
Equations (\ref{rhocrstar-r>a}), (\ref{rhocrstar-r<a}), (\ref{ncinside-star}),
(\ref{veffr>2a-star}), and (\ref{E0-star}) are the main analytical
results for star macroions.

It is important to emphasize that the present approach, while including 
the entropy of the counterions, neglects the configurational entropy 
of the macroions by assuming rigid (fully stretched) PE chains.  
Recently, Jusufi {\it et al}.~\cite{Jusufi} modelled pair interactions 
between PE stars by both molecular dynamics simulation and a variational 
free energy that incorporates chain flexibility. 
An important conclusion of their study is that pair interactions are
dominated by counterion entropy.  Our approach is complementary: while 
the macroion model neglects chain flexibility, which is reasonable 
at least for nonoverlapping stars, the linear response theory refines 
somewhat the modelling of the counterion distribution.

\subsection{Microgel Macroions}\label{ResultsB}
For our model of microgel macroions, we apply exactly the same procedure
as in Sec.~\ref{ResultsA}.
The electric field of a uniformly-charged sphere is
\begin{equation}
E(r)~=~\left\{ \begin{array} {l@{\quad\quad}l}
-\frac{\displaystyle Ze}{\displaystyle \epsilon r^2}, & r>a \\
-\frac{\displaystyle Zer}{\displaystyle \epsilon a^3}, & r\leq a,
\end{array} \right. \label{Egel}
\end{equation}
which integrates to give the macroion-counterion interaction,
\begin{equation}
v_{mc}(r)~=~\left\{ \begin{array} {l@{\quad\quad}l}
-\frac{\displaystyle Zze^2}{\displaystyle \epsilon r}, & r>a \\
-\frac{\displaystyle Zze^2}{\displaystyle 2\epsilon
a}~\left(3-\frac{\displaystyle r^2}{\displaystyle a^2}\right), &
r\leq a.\end{array} \right. \label{vmcrgel}
\end{equation}
Equation (\ref{vmcrgel}) Fourier transforms to
\begin{equation}
\hat v_{mc}(k)~=~-\frac{12\pi Zze^2}{\epsilon
k^4a^2}\left[\cos(ka)-\frac{\sin(ka)}{ka}\right], \label{vmckgel}
\end{equation}
which, when substituted into Eq.~(\ref{rhock}), yields the Fourier transform
of the counterion density profile around a single macroion,
\begin{equation}
\hat\rho_c(k)~=~-\frac{Z}{z}~\frac{3\kappa^2}{k^2a^2(k^2+\kappa^2)}~
\left[\cos(ka)-\frac{\sin(ka)}{ka}\right]. \label{rhockgel}
\end{equation}
The Fourier transform of Eq.~(\ref{rhockgel}) gives the real-space
counterion density profile,
\begin{eqnarray}
\rho_c(r)~&=&~\frac{Z}{z}~\frac{3}{4\pi a^2r}~
\left[\cosh(\kappa a)-\frac{\sinh(\kappa a)}
\kappa a\right]e^{-\kappa r}, \nonumber \\
&&\qquad\qquad\qquad r>a 
\label{rhocrgel-r>a}
\end{eqnarray}
\begin{eqnarray}
\rho_c(r)~&=&~\frac{Z}{z}~\frac{3}{4\pi a^2r}~
\left[\frac{r}{a}-\left(1+\frac{1}{\kappa a}\right)
e^{-\kappa a}\sinh(\kappa r)\right], \nonumber \\
&&\qquad\qquad\qquad r\leq a,
\label{rhocrgel-r<a}
\end{eqnarray}
which approaches a constant as $r\to 0$:
\begin{equation}
\rho_c(r=0)~=~\frac{Z}{z}~\frac{3}{4\pi a^3}~
\left[1-(1+\kappa a)e^{-\kappa a}\right].
\end{equation}

Integration of Eq.~(\ref{rhocrgel-r<a}) yields an analytical result
for the internal fraction of counterions:
\begin{equation}
f_{\rm in}~=~1-\frac{3}{\kappa a} \left(1+\frac{1}{\kappa
a}\right)e^{-\kappa a}\left[\cosh(\kappa a)-\frac{\sinh(\kappa
a)}{\kappa a}\right]. \label{ncinside-gel}
\end{equation}
Again the theory predicts a counterion distribution depending only
on the ratio of macroion radius to screening length and an internal
counterion fraction that increases monotonically with this ratio.  
This prediction may be compared with that of Oosawa's ``two-phase" 
approximation~\cite{PE1}, which assumes uniform (but differing) 
counterion concentrations inside and outside of the macroions. 
According to the latter approach, for spherical macroions with 
volume fraction $\eta$, the condition for equilibrium between 
free and bound counterions, in the absence of salt ions, is
\begin{equation}
\ln\left(\frac{f_{\rm in}}{1-f_{\rm in}}\right)
=\ln\left(\frac{\eta}{1-\eta}\right)
+Z\frac{\lambda_B}{a}(1-f_{\rm in})(1-\eta^{1/3}),
\label{Oosawa}
\end{equation}
which must be solved numerically for $f_{\rm in}$.
The predictions of Eqs.~(\ref{ncinside-gel}) and (\ref{Oosawa})
are compared below in Sec.~\ref{Numerical Results}.

Next, substituting Eq.~(\ref{vmckgel}) into Eq.~(\ref{vindk}),
the induced pair interaction is
\begin{eqnarray}
\hat v_{\rm ind}(k)~&=&~-\frac{36\pi
Z^2e^2}{\epsilon}~\frac{\kappa^2}{k^6a^4(k^2+\kappa^2)} \nonumber \\
~&\times&~\left[\cos(ka)-\frac{\sin(ka)}{ka}\right]^2. \label{vindk-gel}
\end{eqnarray}
For nonoverlapping macroions, the Fourier transform of
Eq.~(\ref{vindk-gel}) is straightforward to evaluate
and yields an effective pair interaction
\begin{eqnarray}
v_{\rm eff}(r)~&=&~
\frac{Z^2e^2}{\epsilon}~\frac{9}{\kappa^4a^4}~\left[\cosh(\kappa
a)-\frac{\sinh(\kappa a)}{\kappa a}\right]^2 \nonumber \\
~&\times&~\frac{e^{-\kappa r}}{r}, \qquad r>2a. \label{veffr>2a-gel}
\end{eqnarray}
As for star macroions, a Yukawa form is predicted, but with a
different amplitude. The case of overlapping macroions is left to
the Appendix.

Finally, from Eqs.~(\ref{E0}), (\ref{vmckgel}), and (\ref{vindk-gel}),
the volume energy is obtained as
\begin{eqnarray}
E_0~&=&~F_{\rm OCP}~-~N_m\frac{3Z^2e^2}{\epsilon
a}\left\{\frac{1}{5}-\frac{1}{2\kappa^2a^2} \right. \nonumber \\
~&+&~\left.\frac{3}{4\kappa^3a^3}
\left[1-\frac{1}{\kappa^2a^2}+\left(1+\frac{2}{\kappa a
}+\frac{1}{\kappa^2 a^2}\right)e^{-2\kappa a}\right]\right\} \nonumber \\
~&-&~(N_+-N_-)\frac{k_BT\kappa^2a^2}{2}. \label{E0-gel}
\end{eqnarray}
Equations (\ref{rhocrgel-r>a}), (\ref{rhocrgel-r<a}), (\ref{ncinside-gel}),
(\ref{veffr>2a-gel}), and (\ref{E0-gel}) are the main analytical
results for microgel macroions.

\section{Numerical Results}\label{Numerical Results}

\begin{figure}
\includegraphics[width=100mm,height=80mm]{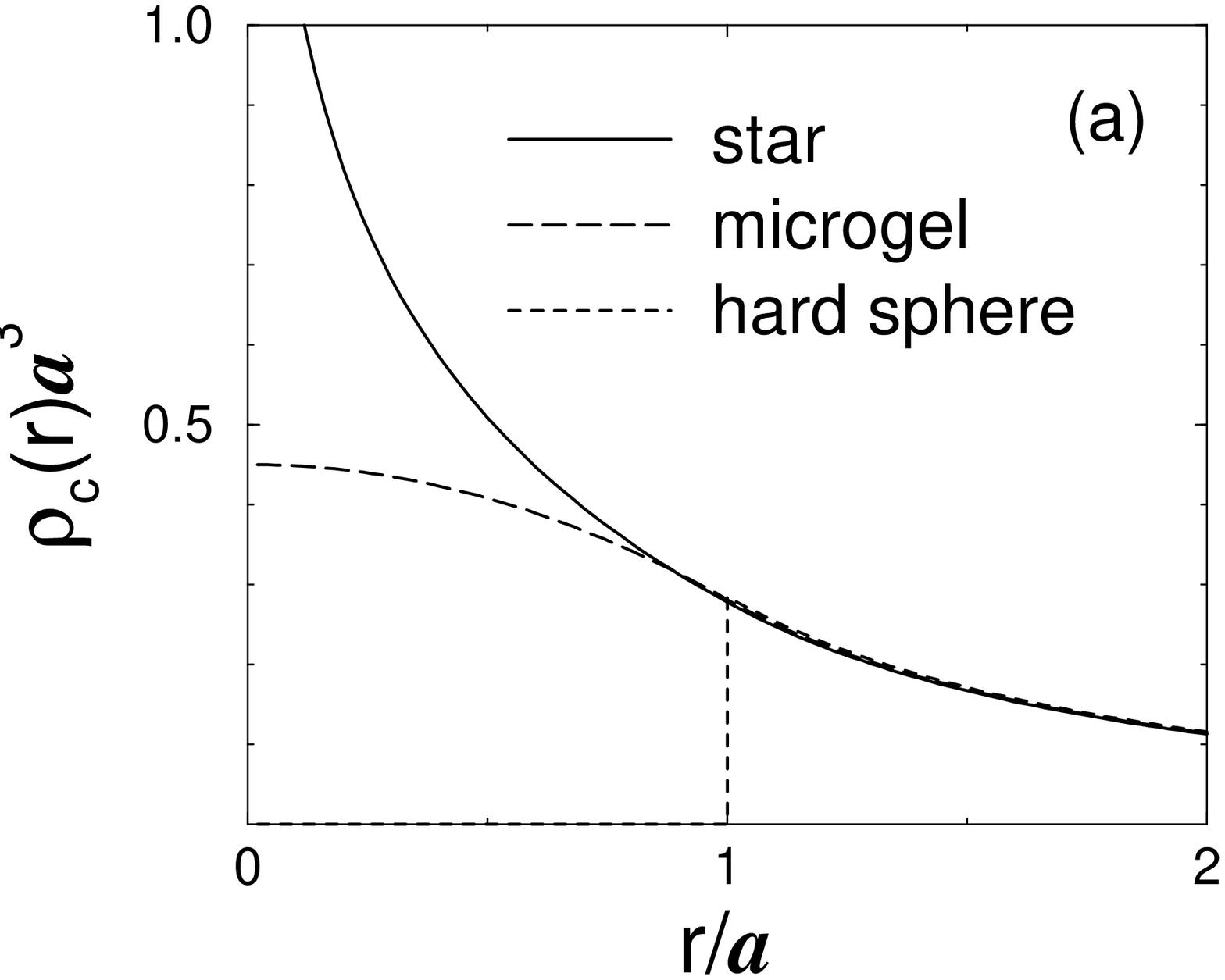}
\includegraphics[width=100mm,height=80mm]{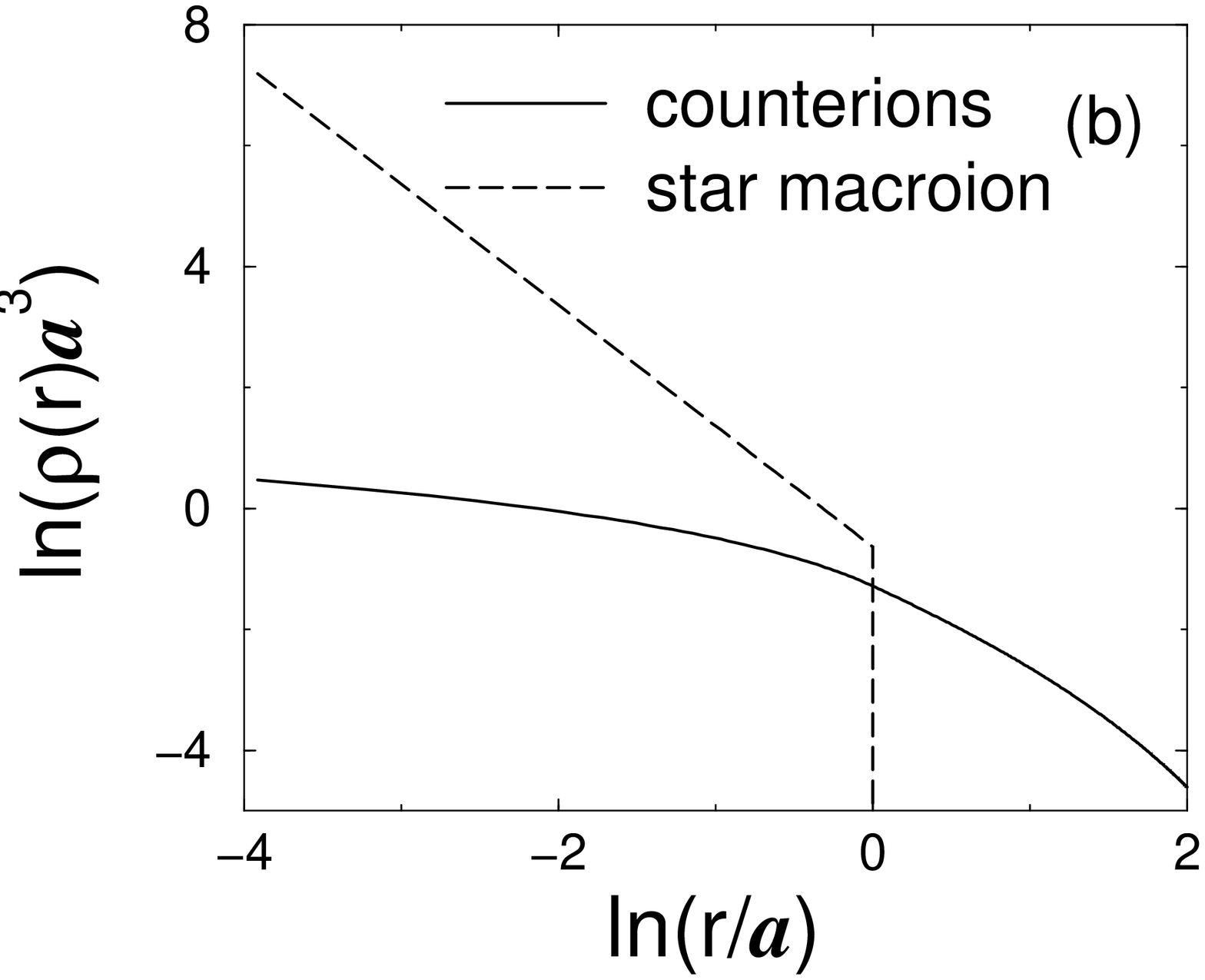}
\caption{\label{rhocstar} (a) Counterion number density profiles 
[from Eqs.~(\ref{rhocrstar-r>a}), (\ref{rhocrstar-r<a}), 
(\ref{rhocrgel-r>a}), and (\ref{rhocrgel-r<a})] 
inside and outside of polyelectrolyte star and microgel macroions of 
diameter $\sigma=100$ nm, valence $Z=100$, and effective volume 
fraction $\eta=0.01$, in water at room temperature ($\lambda_B=0.714$ nm).
The result for a hard-sphere macroion is shown for comparison.
(b) Comparison of counterion and monomer density profiles
for a star macroion on a log-log scale.}
\end{figure}

\begin{figure}
\includegraphics[width=100mm,height=80mm]{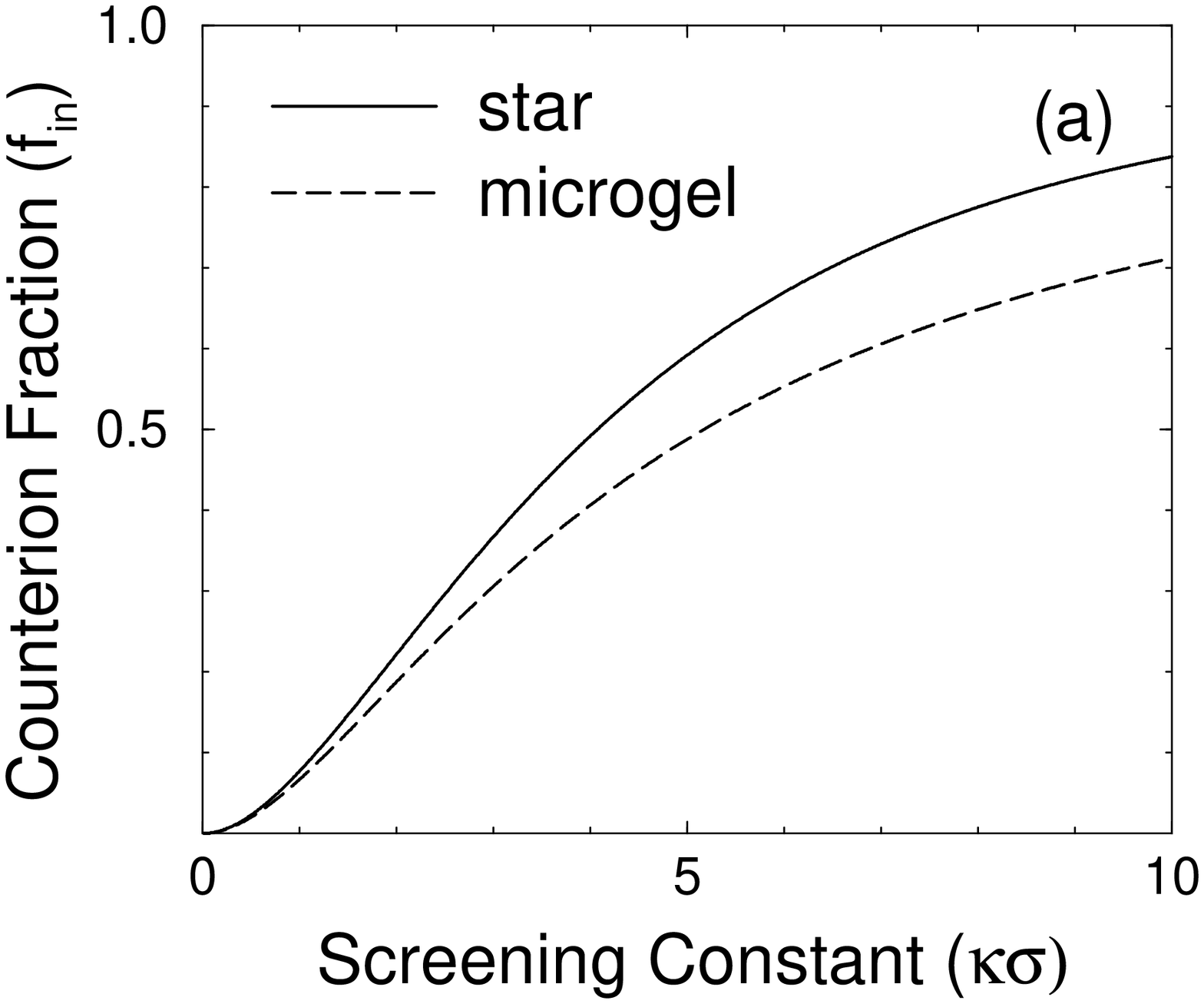}
\includegraphics[width=100mm,height=80mm]{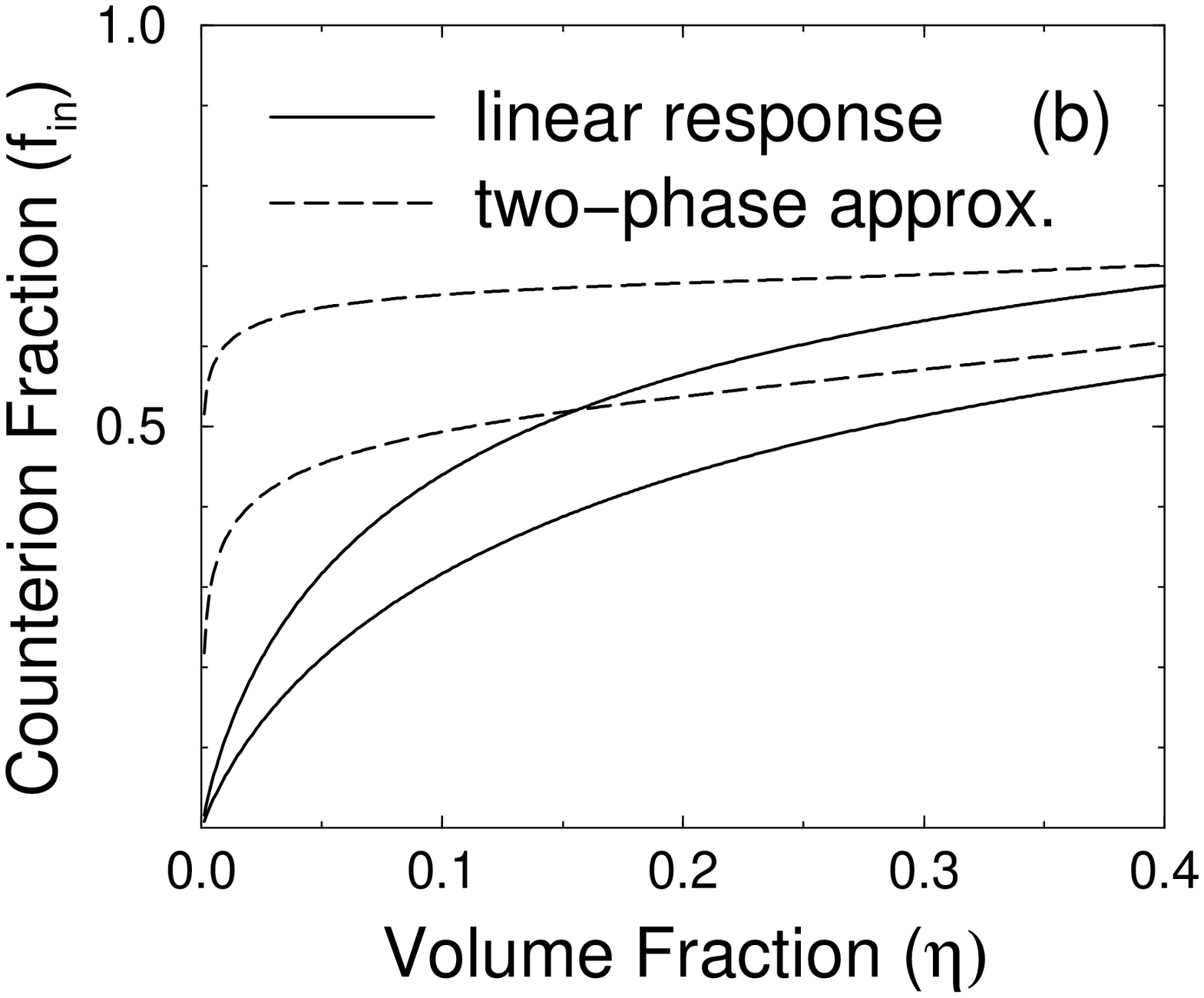}
\caption{\label{ncstar} (a) Fractions of counterions 
[from Eqs.~(\ref{ncinside-star}) and (\ref{ncinside-gel})]
inside polyelectrolyte star and microgel macroions vs.~Debye 
screening constant $\kappa$.  
(b) Fraction of counterions inside a
uniformly charged spherical macroion vs.~effective macroion volume
fraction as predicted by linear response theory (solid curves) and
by the two-phase approximation of Oosawa~\cite{PE1} (dashed
curves). For each case, the bottom curve corresponds to
$Z\lambda_B/a=8$ and the top curve to $Z\lambda_B/a=16$.}
\end{figure}

The theory developed above can be applied to solutions of arbitrary 
ionic strength, under the assumption that the macroion PE chains
remain stretched.  In order to highlight the role of the counterions, 
we present numerical results for salt-free solutions.  
Within the model considered,
the effect of salt is merely to increase the Debye screening constant.
Furthermore, we consider the case of monovalent counterions ($z=1$) 
in aqueous solutions at room temperature ($\lambda_B=0.714$ nm).
Figure \ref{rhocstar} illustrates the form of the 
counterion number density profiles 
[Eqs.~(\ref{rhocrstar-r>a}), (\ref{rhocrstar-r>a}), and
(\ref{rhocrgel-r>a}), (\ref{rhocrgel-r<a})] inside and outside of a macroion. 
Inside a star macroion the counterion density diverges 
logarithmically towards the center, while inside a microgel macroion
$\rho_c(r)$ remains finite.
Evidently, counterions penetrate stars more easily than they do microgels.
This property is also reflected in the internal counterion fractions
[Eqs.~(\ref{ncinside-star}) and (\ref{ncinside-gel})], functions
of $\kappa a$ only, which are shown in Fig.~\ref{ncstar}a. 
In Fig.~\ref{ncstar}b, we compare predictions of linear response theory
[Eq.~(\ref{ncinside-gel})] with those of Oosawa's two-phase
approximation [Eq.~(\ref{Oosawa})] for uniformly-charged spherical 
(microgel) macroions~\cite{PE1}. 
Both approaches qualitatively predict an
increase in the fraction of bound counterions with increasing
macroion concentration. However, linear response theory predicts a
considerably more gradual accumulation of bound counterions than
does the two-phase approximation.

\begin{figure}
\includegraphics[width=100mm,height=80mm]{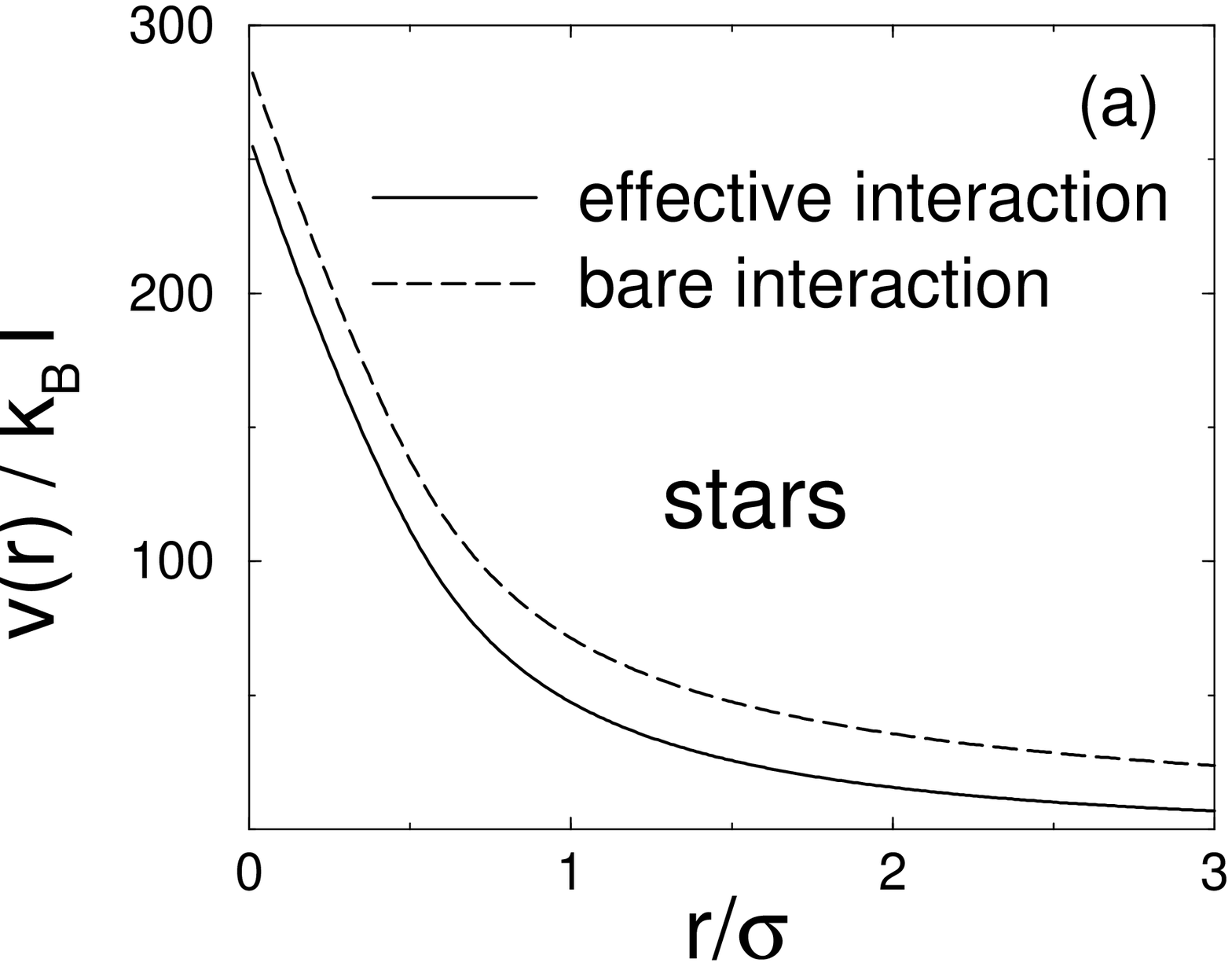}
\includegraphics[width=100mm,height=80mm]{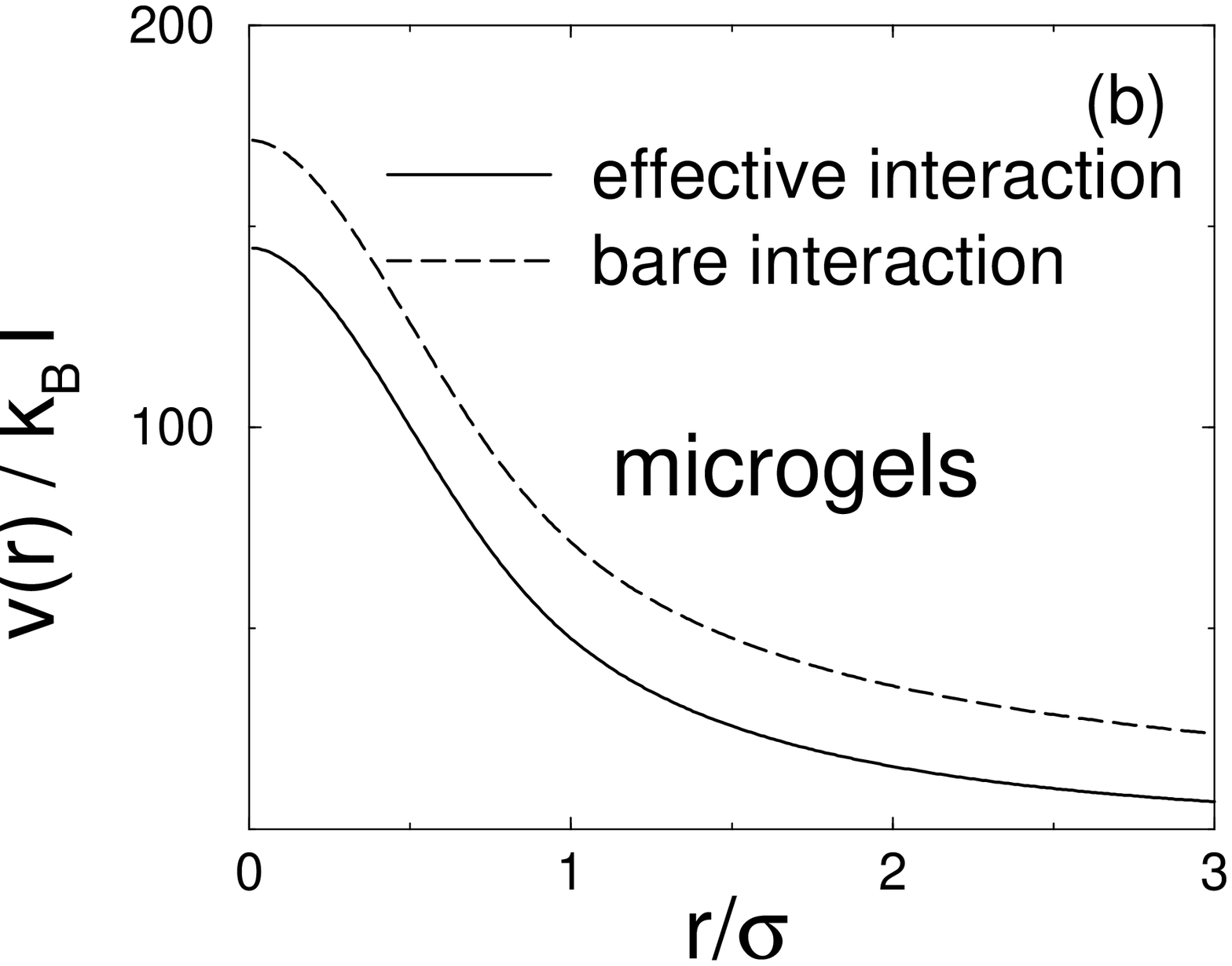}
\caption{\label{vstar} Electrostatic interactions
[from Eqs.~(\ref{veffr>2a-star}) and (\ref{veffr>2a-gel})]
between pairs of polyelectrolyte stars (a) and microgels (b).
Dashed curves: bare interaction.  Solid curves: effective
(bare + induced) interaction.
Parameters are the same as in Fig.~\ref{rhocstar}.
Beyond overlap ($r/\sigma>1$), the interaction is Yukawa in form.
For overlapping macroions ($r/\sigma<1$), the soft repulsion 
remains finite at complete overlap ($r=0$).}
\end{figure}

\begin{figure}
\includegraphics[width=100mm,height=80mm]{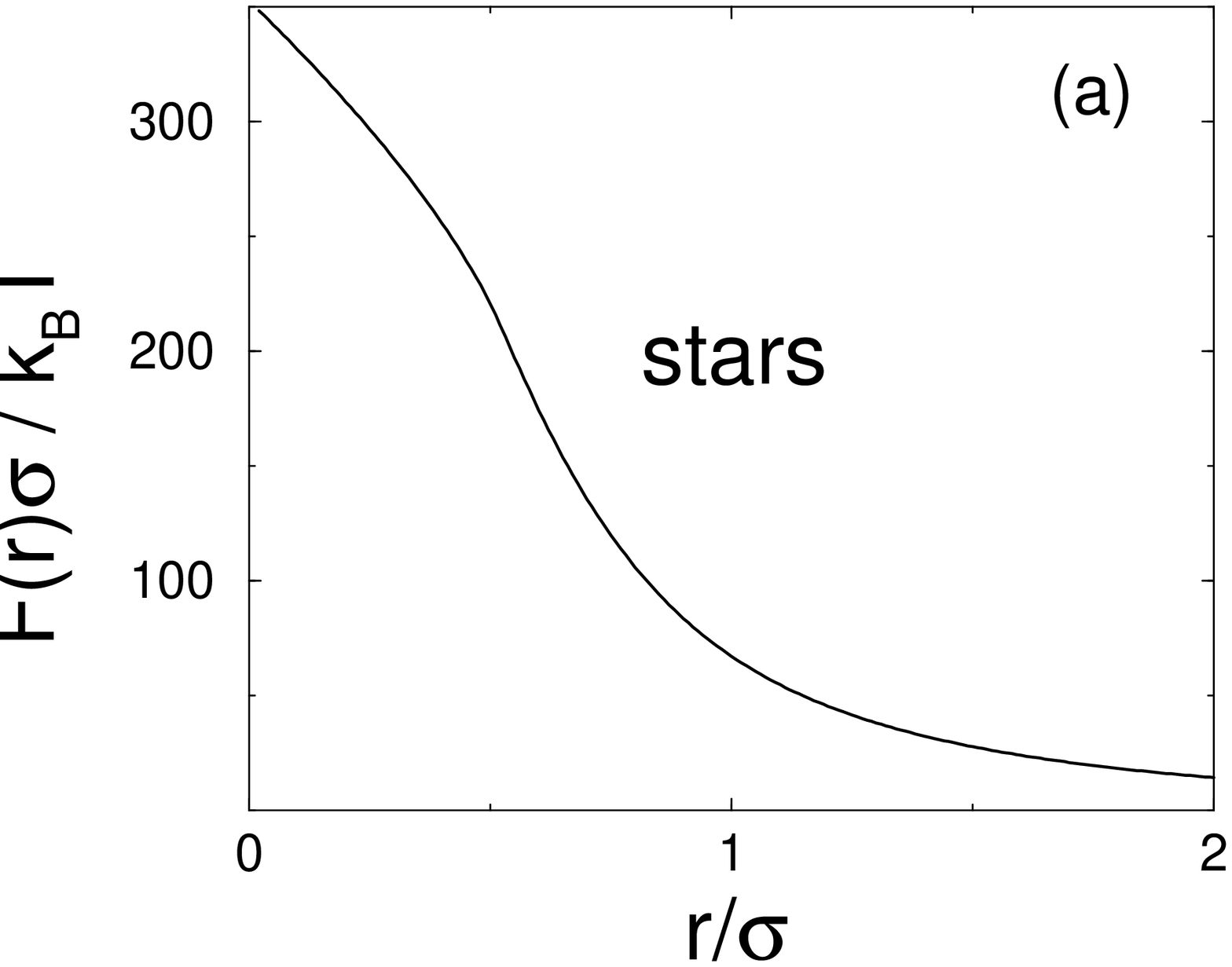}
\includegraphics[width=100mm,height=80mm]{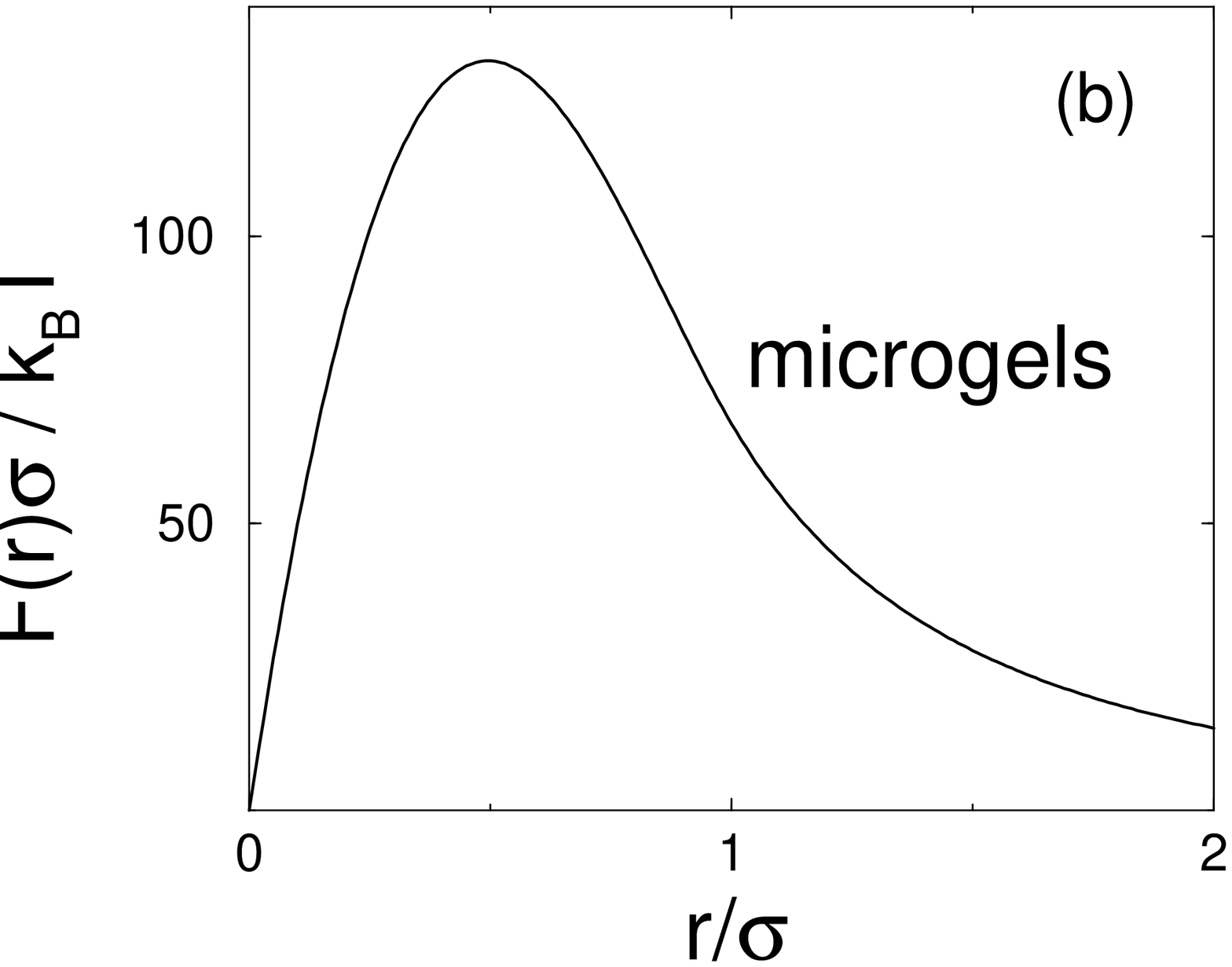}
\caption{\label{fstar} Effective electrostatic forces between
pairs of polyelectrolyte stars (a) and microgels (b),
corresponding to the interaction potentials in Fig.~\ref{vstar}.
Parameters are the same as in Fig.~\ref{rhocstar}.}
\end{figure}

\begin{figure}
\includegraphics[width=100mm,height=80mm]{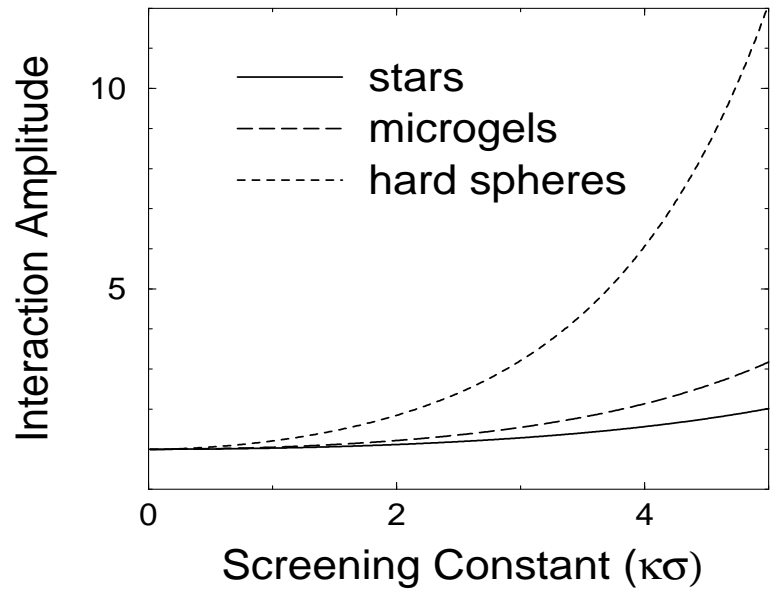}
\caption{\label{vamp} Macroion-size-dependent amplitude of 
Yukawa effective electrostatic interactions 
[Eqs.~(\ref{veffr>2a-star}), (\ref{vdlvo}), and (\ref{veffr>2a-gel})]
between pairs of nonoverlapping stars, microgels, and hard spheres 
vs.~Debye screening constant, normalized to unity at $\kappa\sigma=0$.}
\end{figure}

Counterion penetration strongly influences screening of bare
macroion interactions. The effective pair potentials, $v_{\rm
eff}(r)$, and corresponding forces, $F(r)=-{\rm d}v_{\rm
eff}(r)/{\rm d}r$, are shown in Figs.~\ref{vstar} and \ref{fstar},
respectively. Beyond overlap the effective interaction has Yukawa
form, with the amplitude depending on the type of macroion.
Figure~\ref{vamp} compares the variation of the
macroion-size-dependent amplitude of $v(r>2a)$ with the Debye
screening constant for the two permeable macroions and for
hard macroions.  Evidently, the greater the permeability of the
macroions to counterions, the weaker the amplitude of long-range repulsion.  
For overlapping macroions, the bare charge distribution combined 
with counterion penetration leads to softly repulsive interactions. 
Note that the interactions are bounded: they do not diverge
as the macroions approach complete overlap.
It must be emphasized that the effective interactions presented in
Figs.~\ref{vstar} and \ref{fstar} arise physically from electrostatic 
repulsion and counterion screening, but do not include steric interactions 
due to compression of overlapping chains~\cite{Jusufi}.

\section{Conclusions}\label{Conclusions}

To summarize, we have applied second-order perturbation
(linear response) theory to model solutions of spherical
polyelectrolyte star-branched and microgel macroions. 
The theory predicts the counterion density profiles
inside and outside of the macroions, effective interactions
between pairs of macroions, and a one-body, density-dependent, 
volume energy that contributes to the total free energy of the system. 
The main conclusions are: (1) Counterions penetrate stars more easily 
than they do microgels. (2) Inside a star macroion, the density profile
of mobile counterions varies more gradually than the macroion 
monomer density profile, diverging logarithmically toward the center.  
(3) The fraction of counterions trapped inside a macroion depends 
only on the ratio of macroion radius to Debye screening length
and increases monotonically with this ratio.
(4) Counterion screening significantly weakens the bare electrostatic 
pair interactions, which remain bounded up to complete overlap of macroions.
(5) The effective pair interactions are softly repulsive for 
overlapping macroions and Yukawa in form for separated macroions,
with amplitudes depending on the type of macroion.

It is important to point out some limitations of the
theory. First, the linear response approximation limits
applicability of the theory to dilute solutions of weakly charged
macroions. The quantitative range of validity depends on the
relative magnitudes of nonlinear corrections, including three-body
and high-order interactions, in the perturbation expansion. The
same techniques that have been used to analyze nonlinear response
in charged colloids~\cite{Denton2} can be applied to
polyelectrolytes. Second, the mean-field approach taken here
ignores fluctuations in the counterion distribution, which may be
especially relevant for short-range interactions and multivalent
counterions. Third, the neglect of chain flexibility restricts the
theory to nonoverlapping macroions.  This restriction may be
reasonable for dilute solutions of sparsely separated macroions.
However, for a sufficient concentration of macroions in a good solvent, 
chain elasticity and entropy must play a role. A unification
of linear response theory and the variational theory of 
ref.~\cite{Jusufi} may then prove fruitful.

In principle, the predicted counterion profiles could be probed
experimentally, {\it e.g.}, by neutron scattering, using isotopic
labelling to contrast the PE chains and counterions.
The macroion-macroion interactions may be less
accessible to experiment.  Conceivably, 
the solvent quality might be tuned to minimize the second virial 
coefficient between neutral monomers of overlapping macroions, 
effectively highlighting electrostatic interactions 
by masking any steric interactions.
Comparisons of predicted and observed macroscopic properties 
will provide the most practical, if indirect, tests of the theory.
Future applications will examine thermodynamic phase behavior,
especially possible implications of the volume energy for the
stability and structure of deionized solutions~\cite{microgels}.

\begin{acknowledgments}
It is a pleasure to thank C.~N.~Likos and H.~L\"owen for helpful
discussions and hospitality during a visit to the University
of D\"usseldorf, where parts of this work were completed.
This work was supported by the National Science Foundation
under Grant Nos.~DMR-0204020 and EPS-0132289.
\end{acknowledgments}

\appendix*
\section{Interactions between Overlapping Macroions}

The bare Coulomb interaction between a pair of macroions at
center-center separation $r$ is given in general by
\begin{equation}
v_{mm}(r)~=~\frac{e^2}{\epsilon}\int{\rm d}{\bf r}'\int{\rm d}{\bf
r}'' \,\frac{\rho_{\rm mon}({\bf r}')\rho_{\rm mon}({\bf r}'')}
{|{\bf r}'-{\bf r}''-{\bf r}|}. \label{vppr}
\end{equation}
For nonoverlapping macroions, spherical symmetry reduces the
interaction to
\begin{equation}
v_{mm}(r)~=~\frac{Z^2e^2}{\epsilon r}, \qquad r>2a.
\label{vppr>2a}
\end{equation}
For overlapping macroions, the six-dimensional integral in
Eq.~(\ref{vppr}) may be reduced, by exploiting cylindrical
symmetry and Gauss's law, to two-dimensional integrals, which in
turn may be evaluated analytically. For star macroions, the result
may be expressed piece-wise as follows:
\begin{eqnarray}
&&v_{mm}(r)=\frac{Z^2e^2}{2\epsilon
a}\left\{\frac{9}{2}-\frac{7}{4}\frac{r}{a}-\frac{1}{2}\left[
\left(3-\frac{a}{r}\right)\left(1-\frac{r}{a}\right)\right.\right. \nonumber \\
&+&\left.\frac{r}{a}\ln\left(\frac{r}{a}\right)\right]
\ln\left(\frac{a-r}{a}\right) 
-\frac{r}{2a}\int_{-1}^{a/r-1}{\rm d}x\,\frac{\ln(1+x)}{x} \nonumber \\
&+&\left.\frac{r}{2a}\int_1^{a/r}{\rm d}x\,\frac{\ln(x-1)}{x} \right\}, 
\qquad 0<r\leq a \label{vppr<a1}
\end{eqnarray}

\begin{eqnarray}
&&v_{mm}(r)=\frac{Z^2e^2}{2\epsilon a}\left\{\frac{9}{2}-\frac{7}{4}
\frac{r}{a}-\frac{1}{2}\left[\left(3-\frac{a}{r}\right)
\left(1-\frac{r}{a}\right)\right.\right. \nonumber \\
&+&\left.\frac{r}{a}\ln\left(\frac{r}{a}\right)\right]
\ln\left(\frac{r-a}{a}\right)+\left.\frac{r}{2a}
\int_{1-a/r}^{a/r}{\rm d}x\,\frac{\ln x}{1-x} \right\}, 
\nonumber \\
&&\qquad\qquad\qquad a<r\leq 2a. \label{vbarer<2a1}
\end{eqnarray}

For computational purposes, the remaining
integrals may be expressed as convergent series:
\begin{eqnarray}
&&v_{mm}(r)=\frac{Z^2e^2}{2\epsilon a}\left\{\frac{9}{2}-\frac{7}{4}
\frac{r}{a}-\frac{1}{2}\left[
\left(3-\frac{a}{r}\right)\left(1-\frac{r}{a}\right)\right.\right. 
\nonumber \\
&&\left.+\frac{r}{a}\ln\left(\frac{r}{a}\right)\right]
\ln\left(\frac{a-r}{a}\right)
+\frac{r}{2a}\left(\frac{1}{2}\left[\ln\left(\frac{r}{a}
\right)\right]^2 \right.
\nonumber \\
&&-\frac{1}{2}\left[\ln\left(\frac{r}{a-r}\right)\right]^2
+\sum_{n=1}^{\infty}\frac{(r/a)^n-(r/(a-r))^n}{n^2}
\nonumber \\
&&+\left.\left. 2\sum_{n=1}^{\infty}\frac{(r/(a-r))^{2n-1}-2}{(2n-1)^2}\right)
\right\},~~r\leq a/2 \label{vppr<a/2}
\end{eqnarray}

\begin{eqnarray}
&&v_{mm}(r)=\frac{Z^2e^2}{2\epsilon a}\left\{\frac{9}{2}-\frac{7}{4}
\frac{r}{a}-\frac{1}{2}\left[\left(3-\frac{a}{r}\right)
\left(1-\frac{r}{a}\right) \right.\right.
\nonumber \\
&&\left.+\frac{r}{a}\ln\left(\frac{r}{a}\right)\right]
\ln\left(\frac{a-r}{a}\right)
+\frac{r}{2a}\left(\frac{1}{2}\left[\ln\left(\frac{r}{a}
\right)\right]^2 \right.
\nonumber \\
&&\left.\left.+\sum_{n=1}^{\infty}\frac{(r/a)^n+(1-a/r)^n-2}{n^2}
\right)\right\}, 
\nonumber \\
&&\qquad\qquad a/2<r\leq a \label{vppr<a2}
\end{eqnarray}

\begin{eqnarray}
v_{mm}(r)&=&\frac{Z^2e^2}{2\epsilon a}\left\{\frac{9}{2}-\frac{7}{4}
\frac{r}{a}-\frac{1}{2}\left[\left(3-\frac{a}{r}\right)
\left(1-\frac{r}{a}\right) \right.\right.
\nonumber \\
&+&\left.\frac{r}{a}\ln\left(\frac{r}{a}\right)\right]
\ln\left(\frac{r-a}{a}\right)
\nonumber \\
&-&\left.\frac{r}{2a}\sum_{n=1}^{\infty}\frac{(a/r)^n-(1-a/r)^n}{n^2}\right\}, 
\nonumber \\
&&\qquad\qquad\qquad a<r\leq 2a \label{vppr<2a2}
\end{eqnarray}

For microgel macroions, the bare interaction may be expressed more
compactly as
\begin{eqnarray}
v_{mm}(r)&=&\frac{Z^2e^2}{\epsilon
a}\left[\frac{6}{5}-\frac{1}{2}\left(\frac{r}{a}\right)^2+
\frac{3}{16}\left(\frac{r}{a}\right)^3\right.
\nonumber \\
&-&\left.\frac{1}{160}\left(\frac{r}{a}\right)^5\right], \qquad r\leq 2a.
\label{vppr<2a-gel}
\end{eqnarray}
Finally, the effective pair interaction between microgels,
$v_{\rm eff}(r)$, is the sum of Eq.~(\ref{vppr<2a-gel})
and the induced interaction,
obtained by Fourier transforming Eq.~(\ref{vindk-gel}):
\begin{eqnarray}
v_{\rm
ind}(r)~&=&~-\frac{9Z^2e^2}{2\epsilon\kappa^4a^4r}\left\{\left(
1-e^{-\kappa
r}+\frac{1}{2}\kappa^2r^2+\frac{1}{24}\kappa^4r^4\right)
\left(1-\frac{1}{\kappa^2a^2}\right)
+\frac{2}{\kappa a}e^{-2\kappa a}\sinh(\kappa r) \right.
\nonumber \\
~&+&~\left.\left[e^{-2\kappa a}\sinh(\kappa r)+2\kappa^2ar
+\frac{1}{3}\kappa^4(4a^3r+ar^3)\right]\left(
1+\frac{1}{\kappa^2a^2}\right)
-\frac{2r}{a}\left(1+2\kappa^2a^2+\frac{8}{15}\kappa^4a^4\right)
\right.
\nonumber \\
~&-&\left.~\frac{r^3}{3a^3}\left(\kappa^2a^2+\frac{4}{3}\kappa^4a^4\right)
-\frac{1}{720}\frac{\kappa^4}{a^2}r^6 \right\}, \qquad r\leq 2a.
\label{vindr<2a-gel}
\end{eqnarray}

\clearpage


\begin{references}
\bibitem{note1}
Although the background energy, $E_b$, is strictly infinite in the
idealized case of point counterions, its addition and subtraction
formally cancel corresponding infinities in $H_c$ [Eq.~(\ref{Hc})]
and $H_{mc}$ [Eq.~(\ref{Hmc1})].
\end{references}


\end{document}